\documentstyle[pre,aps]{revtex}

   \font\tenmsb=msbm10 scaled\magstep 1
   \font\sevenmsb=msbm7 scaled \magstep 1
   \font\faivemsb=msbm5 scaled \magstep 1
\newfam\msbfam
      \textfont\msbfam=\tenmsb
      \scriptfont\msbfam=\sevenmsb
      \scriptscriptfont\msbfam=\faivemsb
\def\Bbb#1{{\fam\msbfam #1}}
\font\tengothic=eufm10 scaled\magstep 1
\font\sevengothic=eufm7 scaled\magstep 1
\newfam\gothicfam
      \textfont\gothicfam=\tengothic
      \scriptfont\gothicfam=\sevengothic

\newcommand{\cH}{{\cal H}}
\newcommand{\cL}{{\cal L}}
\newcommand{\cD}{{\cal D}}
\newcommand{\cP}{{\cal P}}
\newcommand{\vp}{\varphi}
\newcommand{\be}{\begin{equation}}
\newcommand{\ee}{\end{equation}}
\newcommand{\ep}{\varepsilon}
\newcommand{\ra}{\rightarrow}
\newcommand{\al}{\alpha}
\newcommand{\bt}{\beta}
\newcommand{\dgr}{\dagger}

\begin{document}

\draft

\title{Entanglement Measure for Composite Systems}
\author{V.I. Yukalov}
\address{Bogolubov Laboratory of Theoretical Physics \\
Joint Institute for Nuclear Research, Dubna 141980, Russia}

\maketitle

\vskip 1cm

\begin{abstract}

A general description of entanglement is suggested as an action realized
by an arbitrary operator over given disentangled states. The related
entanglement measure is defined. Because of its generality, this definition
can be employed for any physical systems, pure or mixed, equilibrium or
nonequilibrium, and characterized by any type of operators, whether these
are statistical operators, field operators, spin operators, or anything
else. Entanglement of any number of parts from their total ensemble forming
a multiparticle composite system can be determined. Interplay between
entanglement and ordering, occurring under phase transitions, is analysed
by invoking the concept of operator order indices.
\end{abstract}

\vskip 1cm


Entanglement is the term Schr\"odinger coined for characterizing superposition
of multipartite quantum states [1], which results in the appearance of specific
quantum correlations between parts of a composite system. Nowadays there is a
growing interest in studying entanglement due to its potential applications in
quantum computing and quantum information processing [2,3]. In order to be a
well defined characteristic, entanglement has to be quantifiable. One usually
considers pairwise entanglement, for which several measures have been suggested,
based on some kinds of reduced or relative entropies [2--8]. In the frame of a
brief communication, it is impossible to present a detailed description of
these measures, whose exhaustive account can be found in books [2,3], reviews
[4--8], and references therein. But it is important to stress that the known
entanglement measures are defined for quantifying only two-partite
entanglement, and there is presently no definitive measure for entanglement
between three or more subsystems. Also, there is no a well defined entanglement
measure for mixed multipartite systems.

The aim of this communication is to introduce a general entanglement
measure that would be valid for any system. Aiming at reaching a high
level of generality, it is necessary, for a while, to leave aside all
physical applications and to focus our attention on the mathematical
structure of the considered concept.

Since entanglement deals with composite systems, we need, first of all,
to concretize the meaning of the latter. In the present case, a system
implies an object characterized by its space of states. Naming a space
composite means that it is composed of some parts. Let the parts be
labelled by an index $i\in\Bbb{I}$. The {\it label manifold} $\Bbb{I}$
can be discrete or continuous. In the simplest case, $i=1,2,\ldots$. Let
each part be characterized by a {\it single-partite space} $\cH_i$, which
is a Hilbert space $\cH_i\equiv\overline{\cL}\{|n_i>\}$, being a closed
linear envelope of a single-partite basis $\{|n_i>\}$. Any vector
$\vp_i\in\cH_i$ is presentable as an expansion $\vp_i=\sum_{n_i} a_{n_i}|n_i>$.
The {\it composite space} $\cH\equiv\otimes_i\cH_i$ is the tensor product,
which is a closed linear envelope $\cH=\overline\cL\{|\{n_i\} >\}$ of
a multipartite basis $\{|\{ n_i\}>\}$ whose vectors can be written as
$|\{ n_i\}>=\otimes_i\;|n_i>$. For any $\vp\in\cH$, one has an expansion
$\vp=\sum_{\{ n_i\} } c_{\{ n_i\} } \; |\{ n_i\}>$.

Two remarks are in order. If the label manifold $\Bbb{I}$ is discrete,
then $\otimes_i\cH_i$ is the standard tensor product with $i=1,2,\ldots$.
When $\Bbb{I}$ is continuous, then $\otimes_i\cH_i$ is the continuous tensor
product, introduced by von Neumann [9] and employed for particular cases in
Refs. [10,11]. Second, $\cH$ is not compulsory a complete tensor product.
It may happen that some selection rules are imposed on the latter, such as
some symmetry requirements. Then $\cH$ is a subspace of the complete tensor
product and it is called [9] incomplete tensor product.

Among all admissible vectors of $\cH$, one may separate out those of two
types. One type forms the {\it disentangled set}
\be
\label{1}
\cD =\{ \otimes_i \vp_i\; | \; \vp_i\in \cH_i \}
\ee
whose vectors $f\in\cD\subset\cH$ have the structure of the tensor product
$f=\otimes_i\sum_{n_i} a_{n_i}|n_i>$ and are termed disentangled states.
All other possible vectors of $\cH$ constitute the complement $\cH\setminus\cD$
whose elements cannot be presented as products of $\vp_i\in\cH_i$ and which are
named entangled states. For instance, in the case of a bipartite system with
two-dimensional single-partite spaces $\cH_i$, the examples of entangled states
would be $c_{12}|12>+c_{21}|21>$ and $c_{11}|11>+c_{22}|22>$, which, clearly,
do not pertain to $\cD$.

Entanglement, by its philological meaning, implies an action or a process by
which disentangled states are transformed into entangled ones. A transforming
action can always be described by an operator. Hence, one may consider
entanglement produced by different operators. Let an operator $A$ be given on
$\cH$. Acting on $\cD$, this operator will, generally, transform disentangled
into entangled states. We shall say that $A$ is an {\it entangling operator},
provided that $A\cD=\cH\setminus\cD$. Of course, not each operator is
entangling. And we shall call $A$ a {\it nonentangling operator} when
$A\cD=\cD$.

For each operator $A$ on $\cH$, we may put into correspondence a nonentangling
operator $A^\otimes$ having the structure of a tensor product $\otimes_i A_1^i$
of {\it single-partite operators}
\be
\label{2}
A_1^i \equiv const\; {\rm Tr}_{ \{ \cH_{j\neq i}\} } A \; .
\ee
In order that the choice of a constant in Eq.(2) would preserve a scale-invariant
form of $A^\otimes$, we require the validity of the normalization condition
\be
\label{3}
{\rm Tr}_{\cH} A = {\rm Tr}_{\cD} A^\otimes \; .
\ee
As a result, we obtain the {\it nonentangling operator}
\be
\label{4}
A^\otimes \equiv \frac{{\rm Tr}_{\cH}A}{{\rm Tr}_\cD\otimes_i A_1^i}\;
\otimes_i A_1^i
\ee
associated with $A$. Note that the trace over the set $\cD$, which is a
restricted subset of the Hilbert space $\cH$, is called the restricted trace.
Such traces are widely used in statistical mechanics [10] and quantum
information theory [2--7]. Their rigorous mathematical definition can be
done by employing the corresponding projecting operators or, more generally,
by invoking weighted Hilbert spaces [10]. In the present case, however, we
do not need a general definition, since here the trace over $\cD$ is applied
only to the factor operators $\otimes_iA_1^i$, for which this reduces just to
the shorthand notation ${\rm Tr}_\cD\otimes_iA_1^i\equiv\prod_i
{\rm Tr}_{\cH_i} A_1^i$.

Now we come to the central problem of quantifying entanglement produced by an
operator $A$ on $\cD$. The principal idea suggesting the way of constructing
this measure stems from the following arguments. Entanglement, generally
speaking, has to do with correlations between parts of a composite system.
Interparticle correlations in physics are often connected with a kind of
order classifying different thermodynamic phases and characterizing phase
transitions. The level of ordering in physical systems can be described by
{\it order indices} [12] advanced for reduced density matrices. This concept
has been generalized by introducing {\it operator order indices} [13] defined
for arbitrary operators. The definition of these order indices involves the
norms of the corresponding operators. It is the operator norm that contains
an essential information on the amount of order hidden in the action of this
operator. Following this way of thinking, the amount of entanglement should
also be related to operator norms. More precisely, we should correlate the
actions on $\cD$ of a given operator $A$ and of its nonentangling counterpart
(4) by comparing the related norms $||A||_\cD$ and $||A^\otimes||_\cD$. Thus
we finally arrive at the definition of the {\it entanglement measure}
\be
\label{5}
\ep(A) \equiv \log\; \frac{||A||_\cD}{||A^\otimes||_\cD} \; ,
\ee
for the entanglement produced by an arbitrary operator $A$ on the disentangled
set $\cD$. As is evident, the measure $\ep(A)=\ep(A,\cD)$ is defined with
respect to $A$ as well as $\cD$. But, for short, we may write $\ep(A)$ when
$\cD$ is fixed. The operator norms can be understood as those associated with
the vector norms, so that $||A||_\cD\equiv\sup_{f\in\cD}||Af||_\cD$, where
$||f||_\cD=1$. The norm over a set $\cD\subset\cH$ is well defined [12],
since it is straightforwardly reformulated to the norm $||A||_\cD\equiv||
\cP_\cD A\cP_\cD||_\cH$ over the Hilbert space $\cH$ by means of the projector
$\cP_\cD$, such that $\cP_\cD\cH=\cD$. Though, in general, it is admissible
to use different kinds of norms, everywhere in what follows the vector norms,
associated with the related scalar products, are employed. This seems to be
more convenient, in particular, because for maximally entangled two dimensional
bipartite states, we get for the entanglement measure $\log 2$. The base
of logarithm may be any, though in information theory it is more customary to
deal with logarithms to the base $2$, when $\log 2=1$. It is easy to show
that the entanglement measure (5) possesses the following natural properties.

\vskip 2mm

1. {\it Semipositivity}: For any bounded operator $A$,
\be
\label{6}
\ep(A) \geq 0 \; .
\ee

\vskip 2mm

2. {\it Nullification}: Measure is zero for nonentangling operators having
the structure of a tensor product $A=A^\otimes$,
\be
\label{7}
\ep(A^\otimes) = 0 \; .
\ee
In particular, there is no self-entanglement of a single part, when
$A=A_1=A^\otimes$, and $\ep(A_1)=0$. The property (7) can be generalized to
the case when $A=\oplus_\nu p_\nu A_\nu^\otimes$, where $||A_\nu^\otimes||_\cD
=||A^\otimes||_\cD$ and $\sum_\nu|p_\nu|=1$, that is,
$\ep\left ( \oplus_\nu p_\nu A_\nu^\otimes\right )=0$.

\vskip 2mm

3. {\it Additivity}: For an operator $A=\otimes_\nu A_\nu$,
\be
\label{8}
\ep(\otimes_\nu A_\nu) = \sum_\nu \ep(A_\nu) \; .
\ee

\vskip 2mm

4. {\it Invariance}: Measure is invariant under local unitary operations
$U_i$,
\be
\label{9}
\ep\left ( \otimes_i U_i^+ A\otimes_i U_i \right ) = \ep(A) \; .
\ee

\vskip 2mm

5. {\it Continuity}: If any considered operator $A$, being parameterized as $A(t)$,
with $t\in\Bbb{R}$, is continuous by norm, such that $||A(t)||_\cD\ra||A(0)||_\cD$
as $t\ra 0$, then measure (5) is also continuous,
\be
\label{10}
\ep(A(t))\ra \ep(A(0)) \qquad (t\ra 0) \; .
\ee

In this way, we may quantify entanglement produced by an arbitrary operator.
Turning to physical systems, we could consider entanglement caused by any
physical operator, for example, due to a Hamiltonian, number-of-particle
operator, momentum, spin, and so on. It is, however, customary to examine
entanglement only with respect to the von Neumann density operator $\hat\rho$.
The entanglement measure (5) can be easily calculated for this operator too,
as we demonstrate below by several examples. For brevity, we shall write $|i>$
instead of $|n_i>$.

\vskip 2mm

(i) {\it Einstein-Podolsky-Rosen states}. The density operator for this famous
example of a pure state is $\hat\rho_{EPR}=|EPR><EPR|$, where $|EPR>\equiv
\frac{1}{\sqrt{2}}(|12>\pm\;|21>)$. In order for readers to clearly understand
how the measure is calculated, we illustrate for the present example all
necessary details. Here the single-partite spaces are $\cH_i$, with $i=1,2$.
Each $\cH_i$ is a two-dimensional Hilbert space, which is a span of the
orthonormalized basis $\{|i>\}$. The composite space is $\cH=\cH_1\otimes\cH_2$.
The disentangled set $\cD$ consists of the product functions $f=\vp_1\otimes
\vp_2$, each $\vp_i$ being a linear combination of the basis vectors $|i>$.
The single-partite operator $\hat\rho_1^i\equiv{\rm Tr}_{\cH_{j\neq i}}
\hat\rho_{EPR}$, following definition (2), becomes $\hat\rho_1=\frac{1}{2}
(|1><1|+|2><2|)$, given on $\cH_i$. The nonentangling operator (4) is
$\hat\rho^\otimes=\hat\rho_1^1\otimes\hat\rho_1^2$. By their definition, all
density operators are self-adjoint. For a self-adjoint operator $\hat\rho$,
the norm, associated with the related vector norm, can be written as
$||\hat\rho||=\sup_{||f||=1}|(f,\hat\rho f)|$. In this way, $||\hat
\rho_1^i||_{\cH_i}=\sup_{j=1,2}<j|\hat\rho_1^i|j>=1/2$. Similarly, $||\hat
\rho_{EPR}||_\cD=\sup_{i,j}<ij|\hat\rho_{EPR}|ij>=1/2$, while for the
nonentangling operator, one has $||\hat\rho^\otimes||_\cD=||\hat\rho_1^1
||_{\cH_1}||\hat\rho_1^2||_{\cH_2}=1/4$. For measure (5), we immediately get
$\ep(\hat\rho_{EPR})=\log 2$, which is unity if the logarithm is to the base
$2$.

\vskip 2mm

(ii) {\it Bell states}. For the density operator $\hat\rho_B=|B><B|$, where
$|B>\equiv\frac{1}{\sqrt{2}}(|11>\pm\;|22>)$, we again have $\ep(\hat\rho_B)=
\log 2$.

\vskip 2mm

(iii) {\it Greenberger-Horne-Zeilinger states}. This is a three-partite state
with $\hat\rho_{GHZ}=|GHZ><GHZ|$, where $|GHZ>\equiv\frac{1}{\sqrt{2}}(|111>
\pm\;|222>)$. The corresponding measure is $\ep(\hat\rho_{GHZ})=2\log 2$.

\vskip 2mm

(iv) {\it Multicat states}. These states are a generalization of the previous
cases, the density operator being $\hat\rho_{MC}=|MC><MC|$, with $|MC>\equiv
c_1|11\ldots 1>+\; c_2 |22\ldots 2>$, where $|c_1|^2+|c_2|^2=1$ and $N$
parts are assumed. The entanglement measure (5) is
$$
\ep(\hat\rho_{MC}) = (1-N)\log\sup\left\{ |c_1|^2,\; |c_2|^2\right\} \; .
$$
The maximum is reached for $|c_1|^2=|c_2|^2=\frac{1}{2}$, when
$\ep(\hat\rho_{MC})=(N-1)\log 2$.

\vskip 2mm

(v) {\it Multimode states}. Such states, that are a generalization of the
multicat states, can be created in coherent systems of $N$ parts, each of
which can accept $m$ different modes [14,15]. For the density operator
$\hat\rho_{MM}=|MM><MM|$, with $|MM>\equiv\sum_n c_n|n\ldots n>$, where
$\sum_n|c_n|^2=1$ and $\sum_n 1=m$, measure (5) becomes
$$
\ep(\hat\rho_{MM}) = (1-N) \log\sup_n |c_n|^2 \; .
$$
Its maximal value happens for $|c_n|^2=1/m$, when $\ep(\hat\rho_{MM})=
(N-1)\log m$.

\vskip 2mm

(vi) {\it Hartree-Fock states}. These describe $N$ parts in different
quantum states. The density operator is $\hat\rho_{HF}=|HF><HF|$, with
$|HF>\equiv\frac{1}{\sqrt{N!}}\sum_{sym} |12\ldots N >$, where a
symmetrized sum is implied, symmetric or antisymmetric, depending on the
type of statistics, Bose or Fermi, respectively. Measure (5) takes the form
$$
\ep(\hat\rho_{HF}) = \log \; \frac{N^N}{N!} \; .
$$

\vskip 2mm

(vii) {\it Mixed states}. The von Neumann density operator can be of any
type, but not only being related to pure states, as in the examples above.
In general, it can be any nonequilibrium operator $\hat\rho(t)$, which leads
to the evolutional entanglement, varying with time [15]. In the case of an
equilibrium system, the von Neumann operator is $\hat\rho=Z^{-1}e^{-\bt H}$,
where $Z$ is a partition function, $\bt$ is inverse temperature, and $H$ is
a Hamiltonian. Measure (5) provides the opportunity of quantifying entanglement
for any $p$ parts of an $N$-partite system. To illustrate this, let us treat
the case when a mixed $p$-partite state is obtained by tracing out $N-p+1$
variables from the Hartree-Fock state of $N$ parts, that is, $\hat\rho^p_{HF}
\equiv{\rm Tr}_{\cH_{p+1}}\ldots{\rm Tr}_{\cH_N}\; \hat\rho_{HF}$. The
entanglement of this $p$-partite mixed state is measured as
$$
\ep\left ( \hat\rho^p_{HF}\right ) = \log \; \frac{(N-p)!N^p}{N!} \; .
$$

\vskip 2mm

(viii) {\it Statistical states}. Mixed states in statistical mechanics, as
is mentioned above, can be characterized by entanglement produced by the von
Neumann statistical operator. However, this is not the sole possibility. Another
way is to measure entanglement realized by reduced density matrices. This way
in many cases may be much simpler, since the properties of the reduced density
matrices have been thoroughly studied [12]. This method also provides a direct
opportunity of quantifying entanglement for any $p$ parts of a given statistical
system. The procedure is as follows. Let $x^p\equiv\{ x_1,x_2,\ldots,x_p\}$ be
a set of variables characterizing the annihilation, $\psi(x)$, and creation,
$\psi^\dgr(x)$, field operators. A $p$-partite density matrix $\rho_p=[\rho_ p
\left (x^p,\overline x^p\right )]$ is defined as a matrix with respect to $x^p$
and $\overline x^p$, with the elements
\be
\label{11}
\rho_p\left ( x^p,\overline x^p\right ) \equiv {\rm Tr}_{\cal F} \psi(x_1)\ldots
\psi(x_p)\hat\rho \psi^\dgr(\overline x_p)\ldots \psi^\dgr(\overline x_1) \; ,
\ee
where the trace is over the Fock space and $\hat\rho$ is a statistical operator.
Under the variables $x$, one may mean, e.g., spatial coordinates, or momentum
variables, or multi-indices in any convenient representation. The studied
statistical system has $N$ parts. These can be indistinguishable particles, with
the field operators satisfying the boson or fermion commutation relations. Then
a $p$-partite density matrix $\rho_p$ describes correlations between any $p$
particles from the ensemble of $N$ identical particles. It is reasonable to
associate the single-partite space $\cH_i$ with a span of the natural orbitals
[12] that are the eigenvectors of the single-partite density matrix
$\rho_1^i=[\rho_1(x_i,\overline x_i)]$. In the considered case, for the
nonentangling operator (4), we have
\be
\label{12}
\rho_p^\otimes = \frac{N!}{(N-p)!\; N^p} \; \otimes_{i=1}^p \; \rho_1^i \; .
\ee
Keeping in mind identical particles, for which $||\rho_1^i||_{\cH_i}=
||\rho_1||_{\cH_1}$, we obtain the entanglement measure (5) as
\be
\label{13}
\ep(\rho_p) = \log \;
\frac{(N-p)! N^p||\rho_p||_\cD}{N!\; ||\rho_1||^p_{\cH_1}} \; .
\ee
What now is left is to find the norms of $\rho_p$ and $\rho_1$ for a given
statistical system, which can be done following the known prescriptions [12].

\vskip 2mm

(ix) {\it Spin states}. Density matrices can be constructed not only of the
field operators, as in Eq. (11), but of any other operators [13]. For instance,
investigating spin systems, one may introduce {\it spin density matrices} [13]
as follows. Let ${\bf S}_i=\{ S_i^\al\}$ be a spin operator associated with
a lattice site $i=1,2,\ldots,N$. We may define a $p$-partite spin density
matrix $R_p=\left [ R_{\{ ij\} }^{\{\al,\bt\}}\right ]$ as a matrix with
respect to all indices, the matrix elements being
\be
\label{14}
R_{\{ ij\} }^{\{\al\bt\}} \equiv {\rm Tr}\; S_{i_1}^{\al_1} \ldots
S_{i_p}^{\al_p} \hat\rho S_{j_p}^{\bt_p} \ldots S_{j_1}^{\bt_1} \; ,
\ee
with the trace over all spin states. The single-partite space $\cH_1$ is defined
as a span of the eigenvectors of the single-partite matrix $R_1=[ R_{ij}^{\al\bt}]$.
The nonentangling matrix (4) is a $p$-fold tensor product $R_p^\otimes=
R_1\otimes R_1\otimes\ldots\otimes R_1$. The entanglement measure (5) becomes
\be
\label{15}
\ep(R_p) = \log\; \frac{||R_p||_\cD}{||R_p^\otimes||_{\cD}} \; .
\ee

\vskip 2mm

(x) {\it Phase transitions}. Since entanglement is related to correlations
existing in a composite system, it would not be surprising if entanglement
would be sensitive to a physical order arising under phase transitions.
Hence the latter can be accompanied by {\it entanglement transitions}.
To prove this, let us consider some examples of phase transitions.

\vskip 2mm

A. {\it Bose-Einstein condensation}. The changes in the reduced density
matrices happening under this transition are well known [12,13]. Essentially
above the condensation point, one has $||\rho_p||_\cD\simeq||\rho_1||^p_{\cH_1}$,
because of which the entanglement measure (13) takes the form $\ep(\rho_p)=
\log (N-p)!N^p/N!$ typical of the mixed Hartree-Fock states. But below the
condensation point, we have $||\rho_p||_\cD=N!/(N-p)!$. Consequently,
entanglement vanishes, $\ep(\rho_p)=0$.

\vskip 2mm

B. {\it Superconducting transition}. Employing the properties of fermion
density matrices, covered in great detail in book [12], we find the following.
Above the critical point, entanglement measure $\ep(\rho_p)$ is again of the
Hartree-Fock form. But below the critical point, one has $||\rho_p||_\cD\simeq
c_pN^{(p-1)/2}$, when $p$ is odd, and $||\rho_p||_\cD\simeq c_pN^{p/2}$, if
$p$ is even [12], where $c_p$ is a constant of order one. Thus, for measure
(13), we obtain
\begin{eqnarray}
\ep(\rho_p) \simeq \left \{ \begin{array}{cc}
\frac{p-1}{2}\; \log N  & (p=1,3,\ldots ) \\
\nonumber
\frac{p}{2}\; \log N & (p=2,4,\ldots ) \; ,\end{array} \right.
\end{eqnarray}
where the large number of particles $N\gg 1$ is assumed. Consequently,
entanglement increases, which is opposite to the case of Bose-Einstein
condensation.

\vskip 2mm

C. {\it Ferromagnetic transition}. We shall study measure (15), based on
spin density matrices [13]. For concreteness, let us keep in mind a Heisenberg
model with long-range interactions, when the mean-field treatment becomes
asymptotically exact in the thermodynamic limit. Then, using the properties
of spin density matrices [13], we can derive measure (15). In the paramagnetic
phase, we have
$$
\ep(R_p) = \log\; \frac{(2p)!}{2^p\; p!} \; ,
$$
but for the ferromagnetic phase we find $\ep(R_p)=0$. Here the situation is
analogous to Bose-Einstein condensation, where the arising long-range order
leads to vanishing entanglement. Such a similarity can be understood if one
remember that, under ferromagnetic phase transition, there occurs condensation
of magnons [13].

As we see, phase transitions are really accompanied by entanglement
transitions. However entanglement behaves differently under different phase
transitions, sometimes vanishing but sometimes increasing, In order to fully
understand the intimate relation between entanglement and ordering, occurring
in physical systems, it is advantageous to resort to the notion of order
indices that have been introduced for density matrices [12] and generalized
to the case of arbitrary operators [13]. The {\it operator order index} for
an operator $A$ is
$$
\omega(A) \equiv \frac{\log||A||}{\log|{\rm Tr}\; A|} \; .
$$
For characterizing ordering in physical systems, the role of $A$ is played
by the appropriate density matrices, such as $\rho_p$ or $R_p$. It is important
that there may develop two types of long-range order, total and even [12,13].
When there is no any order, then $||A||\ll|{\rm Tr}\; A|$ and $\omega(A)\ll 1$.
If, under a phase transition, there develops {\it total order}, then $||A||\sim
|{\rm Tr}\; A|$, so that the order index increases, $\omega(A)\ra 1$, but at
the same time, because of the normalization condition (3), entanglement vanishes,
$\ep(A)\ra 0$. This is the situation taking place at Bose-Einstein condensation
or ferromagnetic transitions. Another case happens under the appearing {\it even
order}, when $||\rho_p||\sim\sqrt{{\rm Tr}\;\rho_p/N}$, if $p$ is odd, while
$||\rho_p||\sim\sqrt{{\rm Tr}\rho_p}$, if $p$ is even. Then the order index
increases to $\omega(\rho_p)=(p-1)/2p$, when $p$ is odd, and to $\omega(\rho_p)
=1/2$, if $p$ is even. This results in the increase of the entanglement measure
$\ep(\rho_p)$, as it is shown above for superconducting transition.

Concluding, a general entanglement measure is introduced, which describes
entanglement realized by an arbitrary operator. For physical systems, it is
convenient to consider entanglement caused by reduced density matrices. The
relation between entanglement measure and order indices is investigated.

\end{document}